# COMPLEMENTARITY RELATIONS FOR MULTI-QUBIT SYSTEMS

Tracey E. Tessier

*Department of Physics and Astronomy*
*University of New Mexico*
*Albuquerque, NM 87131–1156, USA*
*E-mail: tessiert@info.phys.unm.edu*

We derive two complementarity relations that constrain the individual and bipartite properties that may simultaneously exist in a multi-qubit system. The first expression, valid for an arbitrary pure state of $n$ qubits, demonstrates that the degree to which single particle properties are possessed by an individual member of the system is limited by the bipartite entanglement that exists between that qubit and the remainder of the system. This result implies that the phenomenon of entanglement sharing is one specific consequence of complementarity. The second expression, which holds for an arbitrary state of two qubits, pure or mixed, quantifies a tradeoff between the amounts of entanglement, separable uncertainty, and single particle properties that are encoded in the quantum state. The separable uncertainty is a natural measure of our ignorance about the properties possessed by individual subsystems, and may be used to completely characterize the relationship between entanglement and mixedness in two-qubit systems. The two-qubit complementarity relation yields a useful geometric picture in which the root mean square values of local subsystem properties act like coordinates in the space of density matrices, and suggests possible insights into the problem of interpreting quantum mechanics.

## 1. INTRODUCTION

Complementarity is perhaps the most important phenomenon distinguishing systems that are inherently quantum mechanical from those that may accurately be treated classically. Accordingly, a thorough understanding of this concept is of fundamental importance in determining how to properly interpret quantum mechanics [1], as well as in studying how the use of quantum systems might enhance the performance of certain information processing tasks such as communication [2], computation [3], and cryptography [4].

Niels Bohr introduced the term complementarity to refer to the fact that information about a quantum object obtained under different experimental arrangements cannot always be comprehended within a single causal picture [5]. We identify the classical world with precisely those systems and processes for which it is possible to unambiguously combine the space-time coordinates of objects with the dynamical conservation laws that govern their mutual interactions. However, in the more general setting of quantum mechanics, complementarity precludes the existence of such a picture. It was this insight that led Bohr to consider complementarity to be the natural generalization of the classical concept of causality [5].

An alternative statement of complementarity, which makes no reference to experimental arrangements or measurements, states that a quantum system may possess properties that are equally real, but mutually exclusive. The wave-particle duality exhibited by a photon in a double-slit experiment [6] and the tradeoff between the uncertainties in the position and momentum of a subatomic particle governed by Heisenberg's relation [7] are two examples of complementarity in single quantum systems. The study of complementarity in composite systems has a fairly short history by comparison. Nevertheless important progress has been made, especially in the context of two-particle interferometers, where complementarity relations between single and two-particle fringe visibilities [8, 9], between distinguishability and visibility [10], and between the coherence and predictability [11] in a quantum eraser [12] are known and have been experimentally verified [13–15]. Additionally, Jaeger, et al., [16] have recently derived a complementarity relation between multipartite entanglement and mixedness for specific classes of $n$-qubit systems.

Jakob and Bergou [17] took a major step forward by deriving a complementarity relation, valid for an arbitrary pure state of two qubits, which accounts for (and in some cases, generalizes) many of the main results put forward in Refs. [8–11]. They showed that an arbitrary normalized pure state $|\psi\rangle$ of a two-qubit system satisfies the expression [17]

$$C^2 + \nu_k^2 + p_k^2 = 1. \qquad (1)$$

The first term on the left hand side of Eq. (1) is the squared concur-rence [18], a bipartite quantity that, in the context of a system of two qubits, is equivalent to an entanglement monotone [19] known as the tangle [20]. The concurrence $C$ [21], a measure of entanglement in its own right, is defined as

$$C \equiv \left| \langle \psi | \tilde{\psi} \rangle \right|, \tag{2}$$

where $|\tilde{\psi}\rangle \equiv \sigma_y \otimes \sigma_y |\psi^*\rangle$ is the spin-flip [22] of $|\psi\rangle$, and '*' denotes complex conjugation in the standard basis.

The remaining two terms on the left hand side of Eq. (1) are the squares of single particle properties associated with qubit $k = 1, 2$. The first of these properties is the coherence $\nu_k$ of qubit $k$, which quantifies, e.g., the fringe visibility in the context of a two-state system incident on an interferometer. Defining the marginal density operator $\rho_k \equiv \text{Tr}_j(|\psi\rangle \langle \psi|)$ for $j \neq k$, the coherence is given by

$$\nu(\rho_k) \equiv 2 \left| \text{Tr} \left( \rho_k \sigma_+^{(k)} \right) \right|, \tag{3}$$

where $\sigma_+^{(k)} = \begin{pmatrix} 0 & 1 \\ 0 & 0 \end{pmatrix}$ is the raising operator acting on qubit $k$. Similarly, the predictability $p$ which quantifies the *a priori* information regarding whether qubit $k$ is in the state $|0\rangle$ or the state $|1\rangle$, e.g., whether it is more likely to take the upper or lower path in an interferometer, is given by

$$p(\rho_k) \equiv \left| \text{Tr} \left( \rho_k \sigma_z^{(k)} \right) \right|, \tag{4}$$

where $\sigma_z = \begin{pmatrix} 1 & 0 \\ 0 & -1 \end{pmatrix}$ and $|0\rangle$ ($|1\rangle$) is the plus (minus) one eigenvector of $\sigma_z$.

Jakob and Bergou note that Eq. (1) becomes an inequality when applied to a mixed state of two qubits [17]. In this paper we generalize their result in two ways: (i) to apply to an arbitrary pure state of $n$ qubits, and (ii) to apply to an arbitrary state of two qubits, pure or mixed. These expressions shed light on a wide range of topics in quantum information theory, including the highly investigated connection between entanglement and mixedness [23–28] about which it was recently written that, "even for two qubits, the smallest nontrivial bipartite quantum system, the relation between entanglement and mixedness remains a fascinating open question" [28]. Our results lead to a complete characterization of the relationship between these two quantities in exactly this case. Further, they imply that the phenomenon of entanglement sharing, at least in the simplest case of a pure state of three qubits, can be understood as a specific consequence of complementarity.

The remainder of this paper is organized as follows. In Sec. 2. we derive our two generalizations of Eq. (1) from a single common insight and interpret the various quantities involved. One immediate implication of our work is an explicit relationship between the residual tangle [18] and the tangles of the different two-qubit marginals in a pure state of three qubits. The resulting expression illustrates a tradeoff between the various single particle properties, the bipartite entanglement, and the inherent three-body quantum correlations in the tripartite quantum state, effectively reducing the phenomenon of entanglement sharing in this system to a specific instance of complementarity.

Next we derive a quantity, which we dub the separable uncertainty, that arises naturally in the context of our second complementarity relation, and show that it is a good measure of the uncertainty due to ignorance in a quantum state. The introduction of this quantity completes the two qubit picture and enables a comprehensive analysis of the relationship between entanglement and mixedness in these systems. The section ends with several examples designed to illustrate the usefulness of our generalized relations. Finally, we discuss potential applications of our results, as well as some interpretational issues, in Sec. 3..

## 2. MULTI-QUBIT COMPLEMENTARITY RELATIONS

Our two generalizations of Eq. (1) both hinge on the observation (which may be verified by direct computation) that the expression

$$M(\rho_k) = \frac{1}{2} - \overline{S^2}(\rho_k) \qquad (5)$$

holds for an arbitrary state of a single qubit. Here, $M(\rho_k) \equiv 1 - \text{Tr}(\rho_k^2)$ is the marginal mixedness of qubit $k$, and $\overline{S^2}(\rho_k) \equiv 1/2 [\nu^2(\rho_k) + p^2(\rho_k)]$ is the average of the squares of the single qubit properties defined by Eqs. (3) and (4). The mixedness, or linear entropy of a quantum state [29], varies continuously from zero for pure states to its maximum value of $1/2$ for the completely mixed state. The quantity $\overline{S^2}(\rho_k)$ is found to be invariant under local unitary operations performed on qubit $k$, and is therefore taken to be a measure of the single particle properties encoded in $\rho_k$. According to Eq. (5) then, the marginal mixedness quantifies our uncertainty regarding the properties possessed by a single qubit. Further, this expression leads

to the relation

$$\sum_{k=1}^{n} \left[ M\left(\rho_k\right) + \overline{S^2}\left(\rho_k\right) \right] = \frac{n}{2} \quad (6)$$

when summed over an arbitrary system of $n$ qubits, and implies a tradeoff between the single particle properties in such a system and our uncertainty regarding these properties.

Consider now the case where the $n$ qubits are in an overall pure state. The Schmidt decomposition theorem implies that the marginal density operators describing the two subsystems resulting from any bipartite partitioning of this system have the same nonzero eigenvalue spectra [3, 30]. In particular, this holds for $\rho_k$ and $\rho_{\{R_k\}}$, where $\rho_k$ is the marginal state of the $k^{th}$ qubit and $\rho_{\{R_k\}}$ is the marginal state of the $n-1$ qubits in the set $R_k \equiv \{1, 2, \ldots, k-1, k+1, \ldots, n-1, n\}$. Accordingly, a measure of entanglement known as the I-tangle [20], given by $\tau_{k\{R_k\}} = 2M(\rho_k)$, may be used to quantify the quantum correlations that exist between the qubit $k$ and the remaining $n-1$ qubits. Combined with Eq. (6), this yields

$$\sum_{k=1}^{n} \left[ \tau_{k\{R_k\}} + 2\overline{S^2}\left(\rho_k\right) \right] = n. \quad (7)$$

Equation (7) illustrates a complementary relationship between the single particle properties $\overline{S^2}(\rho_k)$ and the $n$ bipartite entanglements $\tau_{k\{R_k\}}$ that may simultaneously exist in an arbitrary pure state of $n$ qubits.

As an interesting application of this expression, consider the case $n=3$. The phenomenon of entanglement sharing was studied in this system by Coffman et. al. [18], where it was realized that the following relationship holds between the amounts of entanglement that may simultaneously exist in different bipartite partitions of the qubits

$$\tau_{1\{23\}} \geq \tau_{12} + \tau_{13}. \quad (8)$$

The general form of the tangle between two qubits, e.g. $\tau_{12}$, is obtained by extending the squared concurrence $C^2$ to mixed states via the convex roof formalism [20]. That is,

$$\tau_{12} = \tau\left(\rho_{12}\right) \equiv \min_{\{p_i, |\psi_i\rangle\}} \sum_i p_i C^2\left(\psi_i\right), \quad (9)$$

where the minimization is performed over all pure state ensemble decompositions of $\rho_{12}$. Equation (8) shows that entanglement cannot be freely distributed among the three qubits, and prompts the definition of the residual tangle [18]

$$\tau_{123} \equiv \tau_{1\{23\}} - \tau_{12} - \tau_{13}, \quad (10)$$

which quantifies the irreducible three-body quantum correlations that exist in the tripartite system. The residual tangle was shown to satisfy the conditions for being an entanglement monotone in Ref. [31].

Using the fact that Eq. (10) is invariant under all permutations of the qubits [18], the residual tangle may be written in the form

$$\tau_{123} = \frac{1}{3}\left\{\tau_{1\{23\}} + \tau_{2\{13\}} + \tau_{3\{12\}} - 2\left[\tau_{12} + \tau_{13} + \tau_{23}\right]\right\}. \quad (11)$$

Expressing Eq. (7) in terms of this symmetrized version of the residual tangle in the case $n = 3$ yields the relation

$$\tau_{123} + \frac{2}{3}\left[\tau_{12} + \tau_{13} + \tau_{23} + \overline{S^2}(\rho_1) + \overline{S^2}(\rho_2) + \overline{S^2}(\rho_3)\right] = 1. \quad (12)$$

It therefore follows that the entanglement sharing behavior exhibited by three qubits in an overall pure state is essential for ensuring consistency with Eq. (12). This expression governs the underlying complementarity that exists between the individual subsystem properties, the bipartite entanglements in the marginal two-qubit states, and the irreducible tripartite quantum correlations. We conjecture that this connection between entanglement sharing and complementarity is a general feature of composite quantum systems.

The derivation of our second generalization of Eq. (1) makes use of a result due to Rungta et al. [32] who showed that (in our notation) $\text{Tr}(\rho\tilde{\rho}) = 1 - \text{Tr}(\rho_1^2) - \text{Tr}(\rho_2^2) + \text{Tr}(\rho^2)$ from which it immediately follows that

$$\text{Tr}(\rho\tilde{\rho}) + M(\rho) = M(\rho_1) + M(\rho_2) \quad (13)$$

for an arbitrary state $\rho$ of two qubits. Here, $\tilde{\rho} \equiv \sigma_y^{\otimes 2}\rho^*\sigma_y^{\otimes 2}$ is the natural generalization of the spin-flip operation to mixed states. According to Eq. (13), $\text{Tr}(\rho\tilde{\rho}) + M(\rho)$ provides an alternative way of calculating the uncertainty regarding single particle properties encoded in such states. Substituting this expression into Eq. (6) in the case $n = 2$ yields

$$\text{Tr}(\rho\tilde{\rho}) + M(\rho) + \overline{S^2}(\rho_1) + \overline{S^2}(\rho_2) = 1. \quad (14)$$

The explicit role played by the sum of the first two terms in Eq. (14) is best illustrated by an example. The invariance under local unitary operations of both $\text{Tr}(\rho\tilde{\rho})$ and $M(\rho)$ allows us to consider, without loss of generality, density matrices $\rho_c$ in the computational basis of the form:

$$\rho_c = \begin{pmatrix} \omega_1 & a & a & e \\ a^* & \omega_2 & f & a \\ a^* & f^* & \omega_3 & a \\ e^* & a^* & a^* & \omega_4 \end{pmatrix}, \quad (15)$$

where $0 \leq \omega_i \leq 1$ and $\sum_i \omega_i = 1$. Equation (15) is obtained by reducing the number of free parameters in an arbitrary density matrix from fifteen to nine via the six free parameters in a tensor product of two single-qubit unitary operators. In this representation, the individual coherences for the two qubits are equal, i.e., $\nu\left(\rho_c^{(k)}\right) = 4|a|$, $k = 1, 2$. Accordingly, $\text{Tr}(\rho\tilde{\rho}) = \text{Tr}(\rho_c\tilde{\rho}_c) = 2\left(|e|^2 + |f|^2 - 4|a|^2 + \omega_2\omega_3 + \omega_1\omega_4\right)$, and $M(\rho) = M(\rho_c) = 1 - 2\left(|e|^2 + |f|^2 + 4|a|^2\right) - \sum_{i=1}^4 \omega_i^2$. Some algebra then yields

$$\text{Tr}(\rho\tilde{\rho}) + M(\rho) = \sum_{i=1}^4 \sigma_i^2 - 2C_{14} - 2C_{23} - \frac{1}{2}\sum_{k=1}^2 \nu^2\left(\rho_c^{(k)}\right), \qquad (16)$$

where $\sigma_i^2 = \omega_i(1 - \omega_i)$ is the variance of $\omega_i$ in the single trial frequencies that result from a measurement in the computational basis, $C_{ij} = -\omega_i\omega_j$ is the similarly defined covariance between $\omega_i$ and $\omega_j$, and $1/2\sum_{k=1}^2 \nu^2\left(\rho_c^{(k)}\right)$ is the average of the squared coherences. The variances measure the spreads or uncertainties associated with the multinomial distribution $\{\omega_i\}$, the covariances are directly related to predictability information that is preserved by the spin-flip operation, and the average squared coherence quantifies the information encoded in the coherences of the individual qubits. Thus, $\text{Tr}(\rho\tilde{\rho}) + M(\rho)$ is the total uncertainty in the distribution $\{\omega_i\}$ minus the available information about properties possessed by the individual subsystems, in complete agreement with Eq. (13). The form of $\rho$ given by Eq. (15) makes this relationship readily apparent; however it holds for an arbitrary density operator due to the invariance of each term in Eq. (14) under local unitary operations.

Equation (14) also enables us to make a connection with the work of Jaeger, et al. [16] who showed that the following expression (in our notation) holds for an arbitrary state of $n$ qubits,

$$\text{Tr}(\rho\tilde{\rho}) + M(\rho) = I(\rho, \tilde{\rho}). \qquad (17)$$

Once again, $\tilde{\rho} \equiv \sigma_y^{\otimes n}\rho^*\sigma_y^{\otimes n}$ is the natural generalization of the spin-flip operation to such systems. The quantity $I(\rho, \tilde{\rho})$, referred to as the indistinguishability, is defined in terms of the Hilbert-Schmidt distance $D_{HS}(\rho - \rho') \equiv \sqrt{\frac{1}{2}\text{Tr}\left[(\rho - \rho')^2\right]}$ between two density matrices $\rho$ and $\rho'$ to be [16]

$$I(\rho, \tilde{\rho}) \equiv 1 - D_{HS}^2(\rho - \tilde{\rho}). \qquad (18)$$

This quantity measures the indistinguishability of the state $\rho$ from the operator $\tilde{\rho}$, and thus serves as a measure of the spin-flip symmetry of the state. Further, Eqs. (13) and (17) imply that, at least in the two-qubit case, the indistinguishability also represents the total uncertainty in the quantum state regarding single particle properties.

Combining Eqs. (14) and (17) in the case $n = 2$ yields

$$I(\rho, \tilde{\rho}) + \overline{S^2}(\rho_1) + \overline{S^2}(\rho_2) = 1, \tag{19}$$

which implies a complementary relationship between information about properties possessed by the individual qubits and the spin-flip symmetry of the state. Substituting Eq. (18) into Eq. (19) then leads to the following relationship between single particle properties and the Hilbert-Schmidt distance between the density operator and its spin-flip,

$$D_{HS}(\rho - \tilde{\rho}) = \sqrt{\overline{S^2}(\rho_1) + \overline{S^2}(\rho_2)}. \tag{20}$$

Equation (20) suggests a geometric picture in which the root mean square values $\sqrt{\overline{S^2}(\rho_k)}$ of the single particle properties act like coordinates in the space of two-qubit density matrices, and shows that the distance between the quantum state and its spin-flip is determined solely by these local properties. Hence, our results yield a method of investigating the abstract space of two-qubit density matrices with simple Euclidean geometry.

Next, in order to determine the specific role played by entanglement in Eq. (14), we make use of the fact that an arbitrary two-qubit density matrix may always be written in its unique optimal Lewenstein-Sanpera decomposition [33]

$$\rho = \lambda \rho_s + (1 - \lambda) |\psi_e\rangle \langle \psi_e|, \tag{21}$$

where $\rho_s = \sum_i p_i \rho_1^{(i)} \otimes \rho_2^{(i)}$, $(0 \leq p_i \leq 1, \sum_i p_i = 1)$ is a separable density matrix, $|\psi_e\rangle$ is an entangled pure state, and $\lambda \in [0, 1]$ is maximal. Calculating the quantity $\text{Tr}(\rho\tilde{\rho})$ using this representation, one finds that

$$\text{Tr}(\rho\tilde{\rho}) = \text{Tr}(\lambda^2 \rho_s \tilde{\rho}_s) + 2\lambda(1-\lambda)\text{Re}\left\langle \tilde{\psi}_e | \rho_s | \tilde{\psi}_e \right\rangle + (1-\lambda)^2 \left|\left\langle \psi | \tilde{\psi} \right\rangle\right|^2. \tag{22}$$

The first term in Eq. (22) quantifies that part of $\rho_s$ which is preserved under the spin-flip operation, and the second quantifies the (real) part of $\left|\tilde{\psi}_e\right\rangle$ that overlaps with $\rho_s$. Hence, neither of these terms involve entanglement. On the other hand, the last term in Eq. (22) is directly related to the quantum correlations in the system as we now demonstrate.

Written in the form of Eq. (21), all of the entanglement in the two-qubit state $\rho$ is concentrated in the pure state $|\psi_e\rangle$ as quantified by the expression [33]

$$C(\rho) = (1-\lambda) C(\psi_e). \tag{23}$$

Since the squared concurrence is equivalent to the tangle for a system of two qubits [20], Eqs. (2) and (23) imply that

$$\tau_{12}(\rho) = C^2(\rho) = (1-\lambda)^2 \left|\langle \psi | \tilde{\psi} \rangle\right|^2, \tag{24}$$

i.e., the last term in Eq. (22) represents the entanglement in the state $\rho$ as quantified by the tangle.

The two fundamental sources of uncertainty regarding single particle properties in composite quantum systems are: (i) ignorance of their values, and (ii) partial to total exclusion of these properties due to the presence of entanglement. Recalling from Eq. (13) that $\text{Tr}(\rho\tilde{\rho}) + M(\rho)$ is a measure of the total uncertainty regarding single particle properties in the quantum state, Eqs. (22) and (24) imply that the quantity

$$\eta(\rho) = \text{Tr}(\rho\tilde{\rho}) + M(\rho) - \tau_{12}(\rho), \tag{25}$$

$0 \leq \eta(\rho) \leq 1$, is a good measure of the *separable uncertainty*, or uncertainty due to ignorance (rather than to the presence of entanglement), in an arbitrary state of two qubits. For example, since $\text{Tr}\left(\psi\tilde{\psi}\right) = \tau_{12}(\psi)$ and $M(\psi) = 0$, we see from Eq. (25) that $\eta(\psi) = 0$, demonstrating that pure states contain no separable uncertainty. Similarly, $\eta(I/4) = 1$ for the completely mixed state, implying that the uncertainty in this case is maximal, and that this state encodes no information regarding either single particle properties or bipartite correlations. Finally, consider the maximally entangled states for fixed marginal mixednesses $\rho_m$ given by [28]

$$\rho_m = \begin{pmatrix} x_1 & 0 & 0 & \sqrt{x_1 x_2} \\ 0 & 0 & 0 & 0 \\ 0 & 0 & 1 - x_1 - x_2 & 0 \\ \sqrt{x_1 x_2} & 0 & 0 & x_2 \end{pmatrix}, \tag{26}$$

with $0 \leq x_1, x_2 \leq 1$ and $x_1 + x_2 \leq 1$. We find that in this case $\eta(\rho_m) = M(\rho_m)$, i.e., the separable uncertainty is simply equal to the mixedness.

Equation (25) completely characterizes the highly investigated connection between entanglement and mixedness in two-qubit systems by relating these quantities to the separable uncertainty and spin-flip invariance encoded in $\rho$. Likewise, combining Eqs. (13) and (25) yields

$$\eta(\rho) = M(\rho_1) + M(\rho_2) - \tau_{12}(\rho). \tag{27}$$

This alternative form of $\eta(\rho)$ quantifies the relationship between entanglement and the marginal mixednesses of the individual qubits.

Our two-qubit generalization of Eq. (1) is finally obtained by combining Eqs. (14) and (25), yielding

$$\eta(\rho) + \tau_{12}(\rho) + \overline{S^2}(\rho_1) + \overline{S^2}(\rho_2) = 1. \tag{28}$$

This expression shows that an arbitrary state of two qubits exhibits a complementary relationship between the amounts of separable uncertainty, entanglement, and information about single particle properties that it encodes. Further, it reduces to Eq. (1) for a pure state $|\psi\rangle$, and has the desirable property that each term is separately invariant under local unitary operations.

The following examples are adapted from Ref. [16] in order to highlight the additional insights provided by Eq. (28) over the previous analysis. Consider the set of states for which $I(\rho, \tilde{\rho}) = 0$, or equivalently, for which $D_{HS}(\rho - \tilde{\rho}) = 1$. These states are maximally distinguishable from their spin-flips. From Eqs. (17) and (25) we see that $\tau_{12}(\rho) = M(\rho) = 0$, which implies that this class of states is equivalent to the set of separable pure states. Equation (28) confirms that in this case $\overline{S^2}(\rho_1) + \overline{S^2}(\rho_2) = 1$, i.e., states of this type possess maximal single particle properties and no bipartite correlations nor separable uncertainty.

Next, consider the class of states for which $M(\rho) = I(\rho, \tilde{\rho})$, i.e., for which the mixedness represents the total uncertainty about single particle properties. Equations (17) and (25) then imply that $\tau_{12}(\rho) + \eta(\rho) = M(\rho)$, which illustrates the precise relationship between entanglement and mixedness for states of this form. Further, from Eqs. (18) and (20) and the fact that the purity $P(\rho) = 1 - M(\rho)$, we surmise that $P(\rho) = \overline{S^2}(\rho_1) + \overline{S^2}(\rho_2)$, yielding an explicit geometric relationship between the purities and the allowable single particle properties of these states.

As a final example, we consider the states that possess perfect spin-flip symmetry. These are the states for which $\rho = \tilde{\rho}$, or by Eq. (18) for which $I(\rho, \tilde{\rho}) = 1$, and hence, $D_{HS}(\rho - \tilde{\rho}) = 0$. Equation (20) then implies that $\overline{S^2}(\rho_1) = \overline{S^2}(\rho_2) = 0$, yielding the result that no state with perfect spin-flip symmetry may encode any information about single particle properties.

A specific class of states satisfying these conditions are the Werner states $\rho_w$ [34]

$$\rho_w(\lambda) = \lambda |\text{Bell}\rangle \langle \text{Bell}| + \frac{1-\lambda}{4} I_2 \otimes I_2, \tag{29}$$

where $0 \leq \lambda \leq 1$, $|\text{Bell}\rangle$ represents one of the four Bell states, and $I_2$ is the identity operator for a single qubit. The Werner states vary continuously from the completely mixed state ($\lambda = 0$) to a maximally entangled state ($\lambda = 1$), and are known to be separable for $\lambda \leq 1/3$ [35]. It is a simple matter to show that the states given by Eq. (29) satisfy the condition that $\rho_w = \tilde{\rho}_w$. Equation (28) then implies that $\eta(\rho_w) = 1 - \tau_{12}(\rho_w)$ demonstrating, among other things, that all separable Werner states are associated with the maximum amount of separable uncertainty, independent of $\lambda$.

Jaeger et al. claim that $\text{Tr}(\rho\tilde{\rho})$ is a good measure of multipartite entanglement, and therefore state that $\text{Tr}(\rho_w \tilde{\rho}_w) + M(\rho_w) = 1$ quantifies the relationship between entanglement and mixedness for the Werner states [16]. However, $\text{Tr}(\rho_w \tilde{\rho}_w)$ fails to satisfy the requirements for being an entanglement monotone [19], since it does not assign the same value to all of the separable Werner states. Indeed our results show that, when considering the class of Werner states the relevant tradeoff occurs not between entanglement and mixedness, but instead between entanglement and separable uncertainty.

## 3. DISCUSSION

The previous examples demonstrate that entanglement and separable uncertainty are quite similar in many respects. For instance, both quantities are related to information in the quantum state (or a lack thereof) which is preserved under the spin-flip operation. Further, both are invariant under local unitary operations, implying that they measure properties which are independent of the choice of local bases. Finally, Eq. (28) shows that both quantities share a complementary relationship with the properties of the local subsystems as well as with one another.

There are also important differences between entanglement and the separable uncertainty quantified by $\eta(\rho)$. First of all, the separable uncertainty vanishes for all pure states, while entanglement is both a pure and mixed state phenomenon. Further, as is well known, entanglement cannot be increased on average by local operations and classical communication (LOCC) [19], while this restriction does not hold for separable uncertainty where we are always allowed to throw away or 'forget' information. Finally, entanglement quantifies the information that we possess regarding the existence of quantum correlations, whereas separable uncertainty quantifies a lack of information about the properties of the individual subsystems.

We conclude from these observations that the entanglement $\tau_{12}(\rho)$, and the single particle properties $\overline{S^2}(\rho_1)$ and $\overline{S^2}(\rho_2)$, are the three fundamental and mutually complementary attributes of a two-qubit system about which we may possess information that does not depend on our choice of local bases. This, in turn, suggests an interpretation for the tangle as the fiducial measure of uncertainty regarding individual subsystem properties due to the presence of entanglement, rather than to our ignorance. Equivalently, because of the relationship between uncertainty and information [36, 37], the tangle also quantifies the amount of information directly encoded in the quantum correlations of the system.

When dealing with quantum systems composed of more than two subsystems, entanglement sharing becomes possible. Equation (12) implies that, at least in the simplest case of a pure state of three qubits, this phenomenon also has its roots in complementarity; this time in terms of a tradeoff between the allowed single particle, bipartite, and tripartite properties that such a system may encode. Verifying the conjecture that entanglement sharing in arbitrary composite systems is generally a consequence of complementarity requires the extension of relations such as Eq. (7) to multipartite systems with subsystems of arbitrary dimension. This in turn requires the identification of the appropriate multipartite generalization of the residual tangle, as well as a determination of which of the possible partitions of such a system contribute to these relations and how to quantify them.

Several potential applications of our second generalized complementarity relation also readily suggest themselves. Beyond fully investigating the relationship between entanglement and mixedness made explicit by Eq. (25), or perhaps more interestingly, between entanglement and the individual subsystem mixednesses given by Eq. (27), our results also seem well-suited to formulating an information vs. disturbance tradeoff relation (see Ref. [38] and references therein) for two-qubit systems. The complementary behavior exhibited by these systems implies that, loosely speaking, a certain amount $\eta(\rho)$ of additional information regarding single particle properties may be obtained through observation without affecting the entanglement in the system. However, if one tries to obtain more information than this, then by Eq. (28) the entanglement must decrease. This behavior leads us to conjecture that complementarity between individual subsystem properties and entanglement plays a fundamental role in the information-disturbance tradeoff phenomenon in composite systems.

Finally, our generalized complementarity relations also suggest one possible way of thinking about the quantum state of a system from an information theoretic point of view. Much has been written about the so-called Bayesian interpretation, which considers the quantum state to be a representation of our subjective knowledge about a quantum system [39]. One advantage of this interpretation is that the collapse of the wave function [40] is viewed not as a real physical process, but simply represents a change in our state of knowledge. However, it is unclear what this knowledge pertains to since, from this point of view, we are generally prohibited from associating objective properties with individual systems. This situation becomes even more confusing if one also contends that a qubit encodes in-principle information i.e. that information is physical [41], since the Bayesian interpretation fails to make a distinction between this type of information and the subjective knowledge of an observer. Equations (7) and (28) provide some insight regarding these observations, especially in the context of two-qubit systems.

We first assume, in agreement with the Bayesian interpretation, that the analysis of any such system must begin with our subjective human knowledge. Accordingly, we assign a quantum state to the system representing this knowledge. Associated with this quantum state assignment is a value for $\eta(\rho)$ which quantifies our subjective uncertainty regarding the in-principle information encoded by the two qubits. In this context Eq. (28) implies that, the smaller our separable uncertainty, the greater our ability to indirectly access and/or manipulate this in-principle information via the locally unitarily invariant bipartite correlations $\tau_{12}(\rho)$ and single particle properties $\overline{S^2}(\rho_k)$, about which we possess subjective information. However, even when we are able to assign a pure state to the quantum system such that $\eta(\psi) = 0$, Eq. (7) substantiates the observation that "maximal information is not complete and cannot be completed" [39]. This is a direct consequence of the complementary relationships that exist between (i) the single particle properties $\nu(\rho_k)$ and $p(\rho_k)$ of the individual subsystems, and (ii) between the total localized attributes $\overline{S^2}(\rho_k)$ of the subsystems and the inherently bipartite entangled correlations $\tau_{k\{R_k\}}$. The logical inconsistencies often associated with quantum mechanics [42] only arise when one mistakenly attributes a greater degree of reality to the individual subsystem properties than is warranted by complementarity.

**Acknowledgements** It is a pleasure to thank Ivan Deutsch, Carl Caves, and Andrew Silberfarb for helpful discussions. This work was partly supported by the National Security Agency (NSA) and the Advanced Research and Development Activity (ARDA) under Army Research Office (ARO) Contract. No. DAAD19-01-1-0648 and by the Office of Naval Research under Contract No. N00014-00-1-0575.